\documentclass[twocolumn,showpacs,preprintnumbers,amsmath,amssymb]{revtex4}

\usepackage{color}
\usepackage{graphicx}
\usepackage{dcolumn}
\usepackage{bm}
\usepackage{array}
\newcommand{\PreserveBackslash}[1]{\let\temp=\\#1\let\\=\temp}
\newcolumntype{C}[1]{>{\PreserveBackslash\centering}p{#1}}
\newcolumntype{R}[1]{>{\PreserveBackslash\raggedleft}p{#1}}
\newcolumntype{L}[1]{>{\PreserveBackslash\raggedright}p{#1}}

\begin{document}

\title{Extracting the information backbone in online system}

\author{Qian-Ming Zhang$^{1}$}
\author{An Zeng$^{2,3}$}
\email{an.zeng@unifr.ch}
\author{Ming-Sheng Shang$^{2}$}
\email{shang.mingsheng@gmail.com}

\affiliation{$^{1}$Web Science Center, School of Computer Science and Engineering,
 University of Electronic Science and Technology of China,\\
Chengdu 611731, People's Republic of China\\
$^{2}$Institute of Information Economy, Alibaba Business College,\\
Hangzhou Normal University, Hangzhou 310036, People's Republic of China\\
$^{3}$Department of Physics, University of Fribourg, Chemin du Mus\'{e}e 3\\
Fribourg CH-1700, Switzerland\\}

\date{\today}

\begin{abstract}
  Information overload is a serious problem in modern society and many solutions such as recommender system have been proposed to filter out irrelevant information. In the literature, researchers mainly dedicated to improve the recommendation performance (accuracy and diversity) of the algorithms while overlooked the influence of topology of the online user-object bipartite networks. In this paper, we find that some information provided by the bipartite networks is not only redundant but also misleading. With such ``less can be more" feature, we design some algorithms to improve the recommendation performance by eliminating some links from the original networks. Moreover, we propose a hybrid method combining the time-aware and topology-aware link removal algorithms to extract the backbone which contains the essential information for the recommender systems. From the practical point of view, our method can improve the performance and reduce the computational time of the recommendation system, thus improve both of their effectiveness and efficiency.
\end{abstract}

\maketitle

\section{Introduction}
Nowadays, we are facing too much information from online systems. We have to make choices from thousands of movies, millions of books, billions of web pages, and so on. The abundant information makes it impossible to go through every candidate products to select the most suitable one. In order to address this problem, many recommendation algorithms have been proposed~\cite{PhysRep2012}. These recommendation systems analyze the purchase history of each user and return with a small number of the most relevant products for him/her. Examples include popularity-based (PR) method, collaborative filtering (CF) method~\cite{CACM4077,ACMTISS225}, mass diffusion (MD) method~\cite{PRE76046115}, heat conduction (HC) method~\cite{PRL99154301}, the hybrid method of mass diffusion and heat conduction~\cite{PNAS1074511} and so on.

The online commercial systems can be represented by the user-object bipartite networks. The recommendation algorithm usually make use of the whole network and the recommendation list is generated based on analyzing all the items bought by the target user~\cite{PhysciaA3911822,EPL9718005}. When the recommendation accuracy is low in some specific online systems, researchers always explain it by the data sparsity~\cite{PhysRep2012}. It is widely believed that the recommendation performance is strongly related to the data amount. However, this common sense might not be true in reality. For instance, when a user bought some items long time ago, these items cannot correctly reflect the current taste of this user. Furthermore, there are always some very popular items, which are almost collected by every user (e.g. some super popular movies watched by everyone). In this case, if a user bought such item, the recommender system cannot extract much information about the user's preference from this purchase action. Therefore, some links in the online user-object bipartite networks can be redundant or even misleading. Appropriately eliminating some connections from the networks might be able to further improve the network function (in our case, recommendation performance). Actually, this ``less can be more'' phenomenon has already been found in many dynamic process. The most well-known example is the synchronization process, in which the synchronizability can be enhanced by removing links \cite{CHAOS037105,NJP14083006}.

The ``less can be more'' feature indicates that there might be backbone structures in the original networks. Generally, a backbone should preserve the topological properties or the function of the original networks. For example, the degree distribution~\cite{PNAS1066483}, betweenness~\cite{PRE70046126}, synchronizability~\cite{PRE73065106,EPL8748002} and transportation ability~\cite{PRL96148702} can be preserved. In online systems, we propose the concept of information backbone which is supposed to preserve the essential information needed for recommendation. By using the information in the backbone structures, the recommender systems are able to make as accurate prediction of users' interested items as the original networks.

In this paper, we consider two main categories of link removal process: time-aware and topology-aware algorithms. We find that both types of algorithms can remove links without significantly harming the recommendation performance. Generally, the time-aware algorithms work better in preserving recommendation accuracy while the topology-aware algorithms have advantage in enhancing the recommendation diversity. We then hybrid these two type of algorithms and achieve a further improvement in preserving the information for recommendation. By using the hybrid algorithm, we obtain the above-mentioned information backbone from the real user-object bipartite networks (The number of links is reduced by about $80\%$). Moreover, the structure properties of the information backbone are analyzed in detail. Finally, we remark that our method is very meaningful from the practical point of view since it can largely reduce the computational cost of the recommender systems.

\section{Methods and Materials}

\subsection{Data Description}
We adopted two standard datasets with time information: Netflix \footnote{http://www.netflix.com} and Movielens \footnote{http://www.movieLens.org}. The Netflix data was sampled from the huge dataset provided for the Netflix Prize. The data is from Feb. 2001 to May 2001 with 8,609 users and 5,081 items. We use the links during the first 3 months as the training set and denote it as $E_{T}$. Among the remaining links, we randomly select some of them as the probe set which is denoted as $E_{P}$. Since the size of $E_P$ cannot be too large compared to $E_{T}$, we set $E_P/E_T\approx 10\%$ in our paper. The training set is treated as known information while the probe set is used for testing and no information in this set is allowed to be used for recommendation. The training set $E_{T}$ of Movielens was sampled from the data collected from unix time 912578016 to 1058210533, i.e. from 2 Dec. 1998 to 15 Jul. 2003. It consists of 5,547 users and 5,850 items. After the unix time 1043723983, the remaining 69,805 links are chosen for the probe set $E_P$ ($E_P/E_T\approx 10\%$). Note that in order to avoid the cold-start problem, we remove all the new users (who rated no items in the training set) and new items (which are not rated by any user in the training set) from the above two probe sets. The simulation is also carried out in other subsets of Netflix and Movielens data and the results are robust, so we only show the result of the above two subsets.

These online commercial systems can be well described by user-object bipartite networks~\cite{EPL8868008}. If a user collects an item, a link is drawn between them. Specifically, we consider a system of $N$ users and $M$ items represented by a bipartite network with adjacency matrix $A$, where the element $a_{i\alpha} = 1$, if a user $i$ has collected an object $\alpha$, and $a_{i\alpha} = 0$, otherwise (throughout this paper we use Greek and Latin letters, respectively, for object- and user-related indices). The aim of the recommender system is to predict which item is most favored by each user, i.e. which element in $A$ is going to change from $0$ to $1$ in the future.

\subsection{Link removal algorithms} \label{partLinkRemv}
In order to examine whether there is abundant (or even misleading) information in the online user-object bipartite
networks, we consider two categories of link removal algorithms: time-aware and topology-aware algorithms.

\emph{time-aware algorithms} use the time information to assign a score for each pair of connected nodes, which is directly defined as their relevance with the underlying assumption that a relevant connection is likely to be a part of the information backbone for recommendation. Here are some typical algorithms:

(1) System oldest removal (SOR): The link appeared earliest among all the remaining links is removed.

(2) System newest removal (SNR): The link appeared latest among all the remaining links is removed.

(3) Individual oldest removal (IOR): The oldest link for each target user is removed.

(4) Individual newest removal (INR): The newest link for each target user is removed.

\emph{topology-aware algorithms} use the network structure to compute the relevance of each link $i\alpha$. Also, we consider four typical algorithms:

(5) Most popular removal (MPR): The popularity of a link $i\alpha$ is defined as $k_{i}k_{\alpha}$, where $k_{i}$ ($k_{\alpha}$) is degree of user $i$ (item $\alpha$). We calculate the popularity of all the remaining links and remove the most popular links.

(6) Least popular removal (LPR): The most unpopular links will be removed.

(7) Most rectangles removal (MRR): A rectangle is defined as a subgraph consisting of four links from two users to two items. We calculate the number of rectangles that each link belongs to, then we remove the link with most rectangles.

(8) Fewest rectangles removal (FRR): We remove the link with fewest rectangles.

Finally, we consider a benchmark algorithm for comparison.

(9) Random removal (RR): Link is randomly chosen and removed.

In order to make all the algorithms comparable, all links should be removed in $50$ macro-steps. Therefore, around $2$ percent links will be chosen in each macro-step. For example, if there are $90$ links in the original network, on average $90/50=1.8$ links should be removed in each macro-step. After $n$th macro-step, $\lceil1.8n\rceil$ links will be removed from the network. In IOR and INR algorithms, the number of links to be removed for each user is proportional to his degree in each macro-step.

\subsection{Recommender system} \label{partRS}
In this paper, we employ the well-known user-based collaborative filtering (UCF) as the standard recommendation system~\cite{CACM4077,ACMTISS225}. In UCF, the recommendation score $f_{\alpha}^{i}$ of an item is evaluated by the similarity $s_{ij}$ between the target user and the users who collected the item,
\begin{equation}
f_{\alpha}^{i}=\sum_{j=1}^{N}s_{ij}a_{j\alpha}.
\end{equation}
Actually, the measure of similarities of two nodes in a network is subject to definition. In this paper, we use the Salton index \cite{Salton} to calculate the similarity between users. For a node $i$, let $\Gamma_i$ denote the set of neighbours of $i$, the Salton index is written as
\begin{equation}
s_{ij}=\frac{|\Gamma_i\bigcap\Gamma_j|}{\sqrt{k_i\times k_j}}
\end{equation}
where $k_i=|\Gamma_i|$ denotes the degree of $i$. The Salton index is also called the cosine similarity in some literatures~\cite{PhysRep2012}.

In this paper, we use several standard metrics to evaluate the recommendation results~\cite{PhysRep2012}. The first one is the area under the receiver operating characteristic (ROC) curve which is used to quantify the accuracy of recommendation~\cite{Hanely_auc}. In the present case, this metric can be interpreted as the probability that a randomly chosen item in $i$'s probe set is given a higher score than a randomly chosen item which is rated by $i$ neither in training set nor in probe set. In the implementation, among $n$ times of independent comparisons, if there are $n'$ times the item in probe set having higher score than the item in the training set and $n''$ times they having the same score, the \emph{accuracy} is defined as:
\begin{equation}
AUC_{i}=\frac{n'+0.5n''}{n}.
\end{equation}

Since real users usually consider only the top part of the recommendation list, a more practical measure may be to consider the number of user $i$'s links in probe set contained in the top $L$ places (It is set as $L=20$ in this paper). This measurement is usually referred as \emph{precision} \cite{Billsus1998} of the recommendation system and the top-$L$ precision is defined as
\begin{equation}
P_{i}(L)=\frac{R_{i}(L)}{L},
\end{equation}
where $R_{i}(L)$ indicates the number of relevant objects (namely the objects collected by $i$ in the probe set) in the top-$L$ places of recommendation list.

Averaging over all the users, we obtain the accuracy and precision of the whole system, as
$AUC=\frac{1}{N}\sum_{i=1}^{N}AUC_{i}$ and $P(L)=\frac{1}{N}\sum_{i=1}^{N}P_{i}(L)$.

Diversity is also an important aspect of recommender system~\cite{PhysRep2012}. Here we adopt inter-user diversity which is defined by considering the uniqueness of different users' recommendation lists. Given two users $i$ and $j$, the difference between their recommendation lists can be measured by Hamming distance,
\begin{equation}
H_{ij}(L)=1-\frac{Q_{ij}(L)}{L},
\end{equation}
where $Q_{ij}(L)$ is the number of common objects in top-$L$ places of both lists. Clearly, if user $i$ and $j$ have the same list, $H_{ij}(L)=0$, while if their lists are completely different, $H_{ij}(L)=1$. Averaging $H_{ij}(L)$ over all pairs of users we obtain the mean distance $H(L)$.

\subsection{Structure indices} \label{partStrucIndex}
After removing links, we will compare the structure features of the obtained network and the original network. The first one is the clustering coefficient~\cite{PRE72056127}, which is defined as the quotient between the number of rectangles and the total number of possible rectangles. For a given node $i$, its clustering coefficient reads
\begin{equation}
C_{4}(i)=\frac{\sum_{m=1}^{k_{i}}\sum_{n=m+1}^{k_{i}}q_{i}(m,n)}{\sum_{m=1}^{k_{i}}\sum_{n=m+1}^{k_{i}}(a_{i}(m,n)+q_{i}(m,n))},
\nonumber
\end{equation}
where $m$ and $n$ label neighbors of node $i$, $q_{i}(m,n)$ are the number of common neighbors between $m$ and $n$ and  $a_{i}(m,n)=(k_m-\eta_i(m,n))(k_n-\eta_i(m,n))$ with $\eta_i(m,n)=1+q_{i}(m,n)$. Here we calculate the the average clustering coefficient of users, items and the whole network respectively. Note that since the nodes whose degrees are below $2$ cannot form any rectangle, we do not take these nodes into account when we calculate the cluster coefficient.

Secondly, we consider the assortative coefficient\cite{Newman2002}, which is the Pearson correlation coefficient of degree between pairs of linked nodes,
\begin{equation}
r=\frac{\sum_{(i,j)\in E}k_{i}k_{j}/|E|-[\sum_{(i,j)\in E}0.5(k_{i}+k_{j})/|E|]^2}{\sum_{(i,j)\in E}0.5(k_{i}^2+k_{j}^2)/|E|-[\sum_{(i,j)\in E}0.5(k_{i}+k_{j})/|E|]^2},
\nonumber
\end{equation}
where $|E|$ is the number of links in a network. Another related index is the degree heterogeneity, calculated on both user side and item side through $H=\langle k^2\rangle/\langle k\rangle^2$.

We also consider the $3$-step diffusion range (DR). It is strongly related to the recommendation process since many recent recommendation algorithms are based on the diffusion process~\cite{PNAS1074511}. For a given node $i$, the $3$-step diffusion range is simply the fraction of covered nodes if the diffusion starts from node $i$ and propagates $3$ steps. The $3$-step diffusion range of a network is the average value of all nodes.

\section{Results}

\subsection{``Less can be more" phenomenon in online systems}

It is usually believed that the more historical information we gather, the more accurate the prediction can be. However, this common sense is not always true, especially in recommender system. In order to examine whether there is abundant (or even misleading) information in the online user-object bipartite networks, we adopted two standard datasets with time information: Netflix and Movielens. We first recall that our main objective is to investigate how much information is needed to correctly predict the links in the probe set and which link removal algorithm is most effective in extracting the essential information from the training set. In our simulation, we will step by step remove links from the training set according to different algorithms (see Subsection \ref{partLinkRemv} ``Link removal algorithms''). After each macro-step, we will monitor the change of the recommendation performance, namely the recommendation accuracy, precision and diversity (see Subsection \ref{partRS} ``Recommender system''). Note that with the macro-step increases, the number of links in the training set gradually decreases while the size of the probe set is always kept unchanged.

\begin{figure*}[!ht]
  \centering
  \includegraphics[width=15cm]{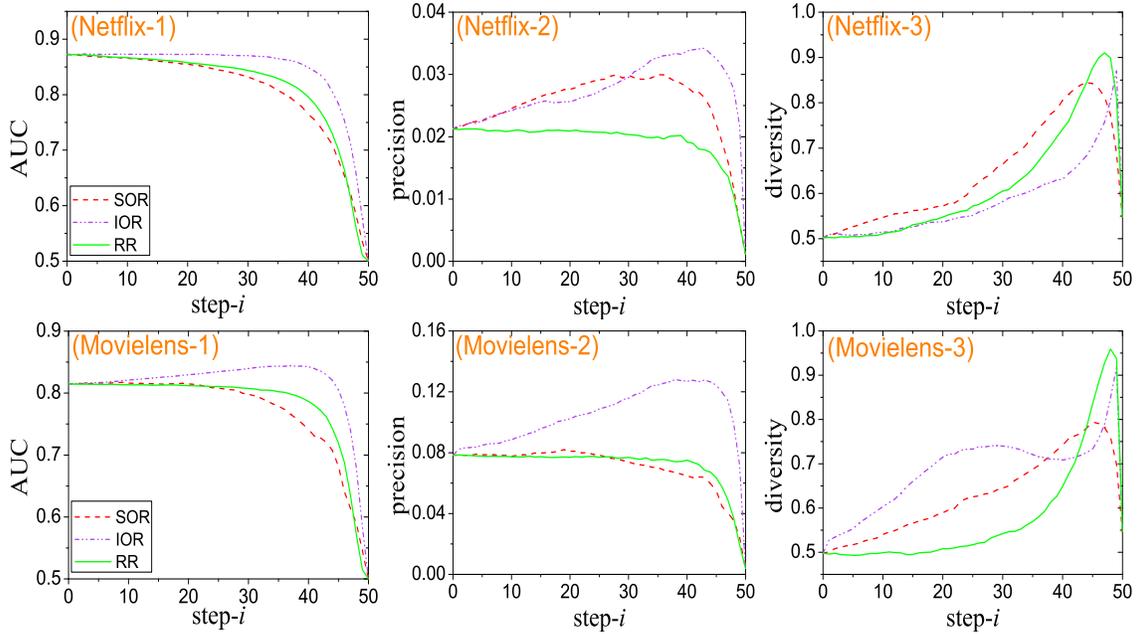}\\
  \caption{
  {\bf The variation tendencies of $AUC$, $P(L)$ and $H(L)$ with the macro-step increases.} step-$i$ is named the identifier of $i$th macro-step. The results of Netflix are shown in sub-figures (Netflix-1), (Netflix-2) and (Netflix-3), and those of Movielens are shown in sub-figures (Movielens-1), (Movielens-2) and (Movielens-3). Note that, only the best performed time-aware algorithms (SOR and IOR) are compared with `Random removal (RR)' here. A comprehensive comparison among these time-aware algorithms is shown in Fig. \ref{fig-Time-Sup} in Appendix SI. }\label{fig-compTime}
\end{figure*}

\begin{figure*}[!ht]
  \centering
  \includegraphics[width=15cm]{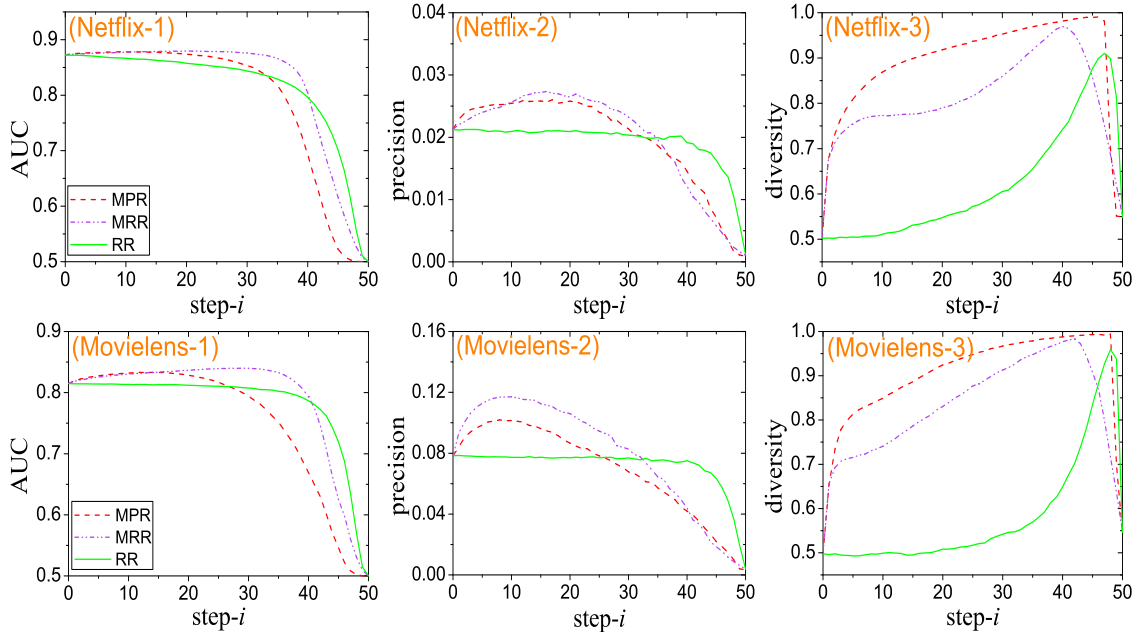}\\
  \caption{
  {\bf The variation tendencies of $AUC$, $P(L)$ and $H(L)$ with the macro-step increases.} step-$i$ is named the identifier of $i$th macro-step. The results of Netflix are shown in sub-figures (Netflix-1), (Netflix-2) and (Netflix-3), and those of Movielens are shown in sub-figures (Movielens-1), (Movielens-2) and (Movielens-3). Note that, only the best performed topology-aware algorithms (MPR and MRR) are compared with `Random removal (RR)' here. A comprehensive comparison among these topology-aware algorithms is shown in Fig. \ref{fig-Topo-Sup} in Appendix SI.} \label{fig-compTopo}
\end{figure*}

The results for the time-aware algorithms are reported in Fig. \ref{fig-compTime} (note that only the most related results are plotted here for the sake of clear presentation and the comprehensive comparison is shown in Fig. \ref{fig-Time-Sup} in Appendix SI). Interestingly, instead of decreasing, the $AUC$ and $P(L)$ can increase as the links are removed from the network based on some algorithms. Overall speaking, SOR and IOR perform better in time-aware algorithms, while the recommendation accuracies of the other two, i.e., SNR and INR, decline sharply. Many studies have revealed that putting less weight on the old links can indeed improve the recommendation performance~\cite{PhysicaA393643}. Therefore, SOR and IOR work well in the link removal process. In our simulation, we observe that IOR is generally better than SOR. This is because SOR may remove all links for some small degree users, which leads to very serious cold-start problem.

The results for the topology-aware algorithms are reported in Fig. \ref{fig-compTopo} (again only the most related results are plotted for the sake of clear presentation and the comprehensive comparison is shown in Fig. \ref{fig-Topo-Sup} in Appendix SI). In the topology-aware algorithms, the MPR and MRR are more accurate than others. In the previous literatures, it shows that the recommendation performance is strongly related to the clustering effect of the networks~\cite{MS531146}. More specifically, the more rectangles the network has, the more accurate the recommendation can be. In this sense, the link with few rectangles do not have much information and should be removed first from the network. However, we show that MRR algorithm performs far better than the FRR. Similar phenomenon is observed in the algorithms which consider the link popularity. In the item side, the most popular items are bought by almost all the users. The links connecting to the hub items cannot reflect the real taste of users. Likewise, a high degree users are interested in many different kinds of items. If an item is collected by such user, the recommendation system cannot determine the intrinsic property of this item and thus cannot predict the potential users who might like it. Therefore, the links with low popularity generally contain more information. Moreover, the MPR and MRR algorithms not only help the recommendation system to reveal the real taste of users, but also improve the recommendation diversity (see Fig. \ref{fig-compTopo}).

In both Fig. \ref{fig-compTime} and Fig. \ref{fig-compTopo}, we plotted the results of random removal (RR) for comparison. It seems that the recommendation accuracy can be also well preserved in RR algorithm. However, RR cannot improve the AUC and precision by removing links as the SOR, IOR, MPR and MRR algorithms. Besides, the recommendation diversity is very low when using the RR algorithm. Since the links of the small degree users and unpopular items have the same probability as the other links to be removed, the RR algorithm will cause quite serious cold-start problem.

The phenomenon above indicates that there is ``less can be more'' feature in the online recommendation system. At the beginning, some redundant and misleading links are deleted, which improves the recommendation accuracy and precision. As links are removed, some necessary information for the recommender systems will be inevitably destroyed, and thus both the accuracy and precision decrease in the final part of link removal process as shown in Fig. \ref{fig-compTime}. These results imply that there is an information backbone of these online bipartite networks.

\subsection{The information backbone and the related topology properties}

By comparing the performances of different removal algorithms, we find that both the time-aware algorithms and topology-aware algorithms can remove the redundant and misleading information from the networks. However, each type of methods has its own advantage. The time-aware algorithms work better in preserving recommendation accuracy while the topology-aware algorithms have advantage in enhancing the recommendation diversity. One very straight forward extension is to hybrid the methods to better extract the information backbone from the online bipartite networks. For simplification, we chose SOR in the time-aware algorithms and MPR in the topology-aware algorithms. We use a tunable parameter $\lambda$ in the hybrid method to adjust the tendency for the SOR algorithm and MPR algorithm. In practice, a random number $N_{rand}$ between $0$ and $1$ is generated before removing a link. If $N_{rand}>\lambda$, the link should be selected according to SOR; or else, it should be selected according to MPR.

\begin{figure*}[!ht]
  \centering
  \includegraphics[width=15cm]{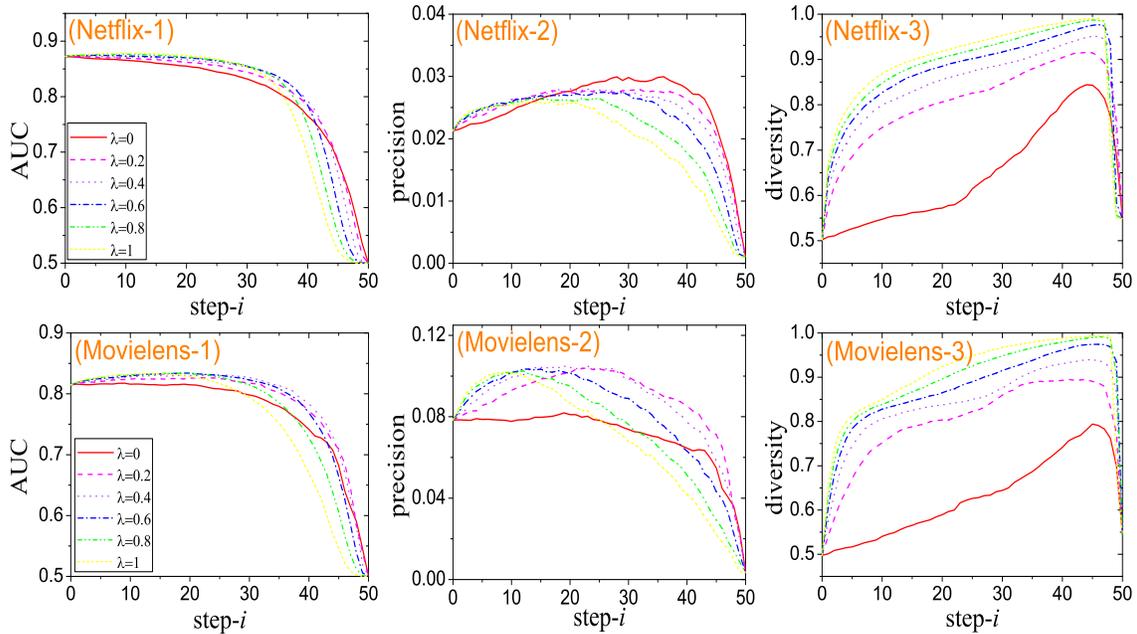}\\
  \caption{
  {\bf The dependence of accuracies and diversities on $\lambda$.} Sub-figures (Netflix-1), (Netflix-2) and (Netflix-3) are corresponding to Netflix and other sub-figures are corresponding to Movielens.}\label{fig-hybrid}
\end{figure*}

The results of this hybrid method are shown in Fig. \ref{fig-hybrid}. When $\lambda=0$ (pure time-aware algorithm), although the recommendation accuracy and precision can stay relatively high even a lot of links are removed, the recommendation diversity is not satisfying enough. When $\lambda=1$ (pure topology-aware algorithm), the recommendation diversity can be very close to the maximum $1$. However, the recommendation accuracy and precision drop quickly as the links are removed. The hybrid algorithm is able to keep a reasonable balance between recommendation diversity and accuracy. Moreover, the hybrid algorithm can sometimes even outperform the time-aware algorithm in preserving the recommendation accuracy when a large number of links are removed from the networks.

With the hybrid method, we further move to extract the information backbone from the bipartite networks. One immediate question is how many links should be removed. Here, we use a simple criteria to determine the optimal number of links to remove. As discussed above, the backbone should effectively preserve the recommendation accuracy of the original networks. In the hybrid method, links are removed until the $AUC$ is lower than $95\%$ of the $AUC$ of the original networks. We select the $\lambda$ under which the number of removed links are the largest. Note that, when there are several $\lambda$ with the same number of removed links, we select the one with the highest recommendation diversity. In the way, we can get the information backbone of the original networks. In this backbone, the recommendation performance is preserved and the recommender systems only have to deal with a small number of links (72\% and 80\% links are removed in Movielens and Netflix, respectively). The related results can be seen in Table \ref{TableComparison}. It shows that the resulting network from the hybrid algorithm has both high recommendation accuracy and diversity compared to the pure algorithms.

\begin{table*}[!ht]
  \caption{\bf {Comparisons of the results among initial network and the resulting networks by different algorithms.} }
  \centering
  {\begin{tabular}{l|cccc|cccc}
  \hline
  \hline
   & \multicolumn{4}{c|}{Netflix} & \multicolumn{4}{c}{Movielens} \\
   & InitialData & SOR$^{36}$ & MPR$^{36}$ & Hybrid$_{0.6}^{36}$ & InitialData & SOR$^{40}$ & MPR$^{40}$ & Hybrid$_{0.5}^{40}$ \\
  \hline
  $AUC$             &0.8725 &0.8013 &0.8012 &0.8300 &0.8148 &0.7413 &0.6692 &0.7831\\
  $P(L)$            &0.0213 &0.0299 &0.0178 &0.0243 &0.0783 &0.0649 &0.0419 &0.0700\\
  $H(L)$            &0.0519 &0.7441 &0.9718 &0.9394 &0.4976 &0.7418 &0.9877 &0.9498\\
  $C4_{user}$       &0.0008 &0.0019 &0.0006 &0.0007 &0.0009 &0.0036 &0.0007 &0.0011\\
  $C4_{item}$       &0.0022 &0.0024 &0.0015 &0.0013 &0.0026 &0.0012 &0.0007 &0.0006\\
  $C4_{net}$    &0.0013 &0.0021 &0.0010 &0.0010 &0.0017 &0.0018 &0.0007 &0.0008\\
  $r$               &0.0470 &0.0312 &0.4922 &0.4839 &0.4131 &0.3331 &0.9459 &0.6349\\
  $H_{user}$        &4.04   &10.22  &1.57   &2.34   &3.43   &14.05  &1.16   &2.82  \\
  $H_{item}$        &6.99   &8.94   &2.37   &3.51   &3.74   &4.21   &1.28   &1.75  \\
  $DR$      &0.6418 &0.1967 &0.1867 &0.1911 &0.9092 &0.2347 &0.3165 &0.2584  \\
  \hline
  \hline
  \end{tabular}
  \begin{flushleft}In ``SOR$^{b}$'', ``MPR$^{b}$'' and ``Hybrid$_{a}^{b}$'', $a$ is the optimal $\lambda$ and $b$ is the identifier of the corresponding macro-step of backbone.
  \end{flushleft}
  \label{TableComparison}}
\end{table*}

Next, we try to investigate the structure features of the obtained information backbone. We compare the original networks and the obtained information backbone in four structure indices here: clustering coefficient, assortativity, degree heterogeneity and $3$-step diffusion range (See subsection \ref{partStrucIndex} ``Structure indices''). The structural properties of the initial network and the resulting networks by different algorithms can be also seen in Table \ref{TableComparison}. Clearly, the structure properties of the network from the hybrid algorithm (which we call ``information backbone") is between the SOR and MPR algorithms. The clustering coefficient of the information backbone is inevitably smaller than the original networks since clustering coefficient is strongly related to the link sparsity. For the assortativity, the information backbone generally has higher value than the original networks. As mentioned above, the links to the hubs items cannot reflect the real interests of the users, so these links are removed from the networks. Therefore, a lot of links connecting to hub items and hub users are removed. As a result, the assortativity is generally larger in the backbone networks and this also explains why the degree heterogeneity of the backbone network is generally smaller. As for the $3$-step diffusion range, the information backbone contains essential information for recommendation system. The items reached by $3$-step diffusion are almost all the items which might be interested by the users. The wrong items are no longer covered by the diffusion. Therefore, the diffusion range is much smaller than the original networks.

\section{Discussion}
The rapid expansion of the internet leads to an increasing amount of information from the World Wide Web. Recommendation algorithms are thus proposed to address the problem of information overload. Previous recommendation algorithms use all the available information of the online user-object bipartite networks to generate the recommendation list. We find, however, that some links in the networks might be redundant and misleading. Therefore, we proposed a hybrid algorithm combining both the time and topology information to remove unnecessary links. In this way, we obtained the information backbone which contains the essential information for recommendation.

Nowadays, the recommendation systems have to deal with very large amount of data to generate personalized recommendation for each user. Actually, the backbone extraction method can be regarded as the data pretreatment. Before the recommendation is implemented, the amount of data can be significantly reduced by our method while the recommendation results can stay almost the unchanged. In this sense, our method can be very meaningful in practical point of view since it can largely reduce the computational cost of the recommendation systems.

\section*{Acknowledgments}
We thank Tao Zhou for helpful discussion. This work is supported by the opening foundation of Institute of Information Economy in Hangzhou Normal University (Grant No. PD12001003002002) and National Natural
Science Foundation of China (Grants No. 61073099, No. 60973069, No. 60973069). QMZ acknowledges the supporting from the Program of Outstanding PhD Candidate in Academic Research by UESTC (Grant No. YBXSZC20131034).

\section*{Supporting Information}
\textbf{Appendix SI} Appendix to the manuscript.\\
(PDF)

\setcounter{figure}{0}
\renewcommand\thefigure{S\arabic{figure}}
\begin{figure*}[!ht]
  \centering
  \includegraphics[width=15cm]{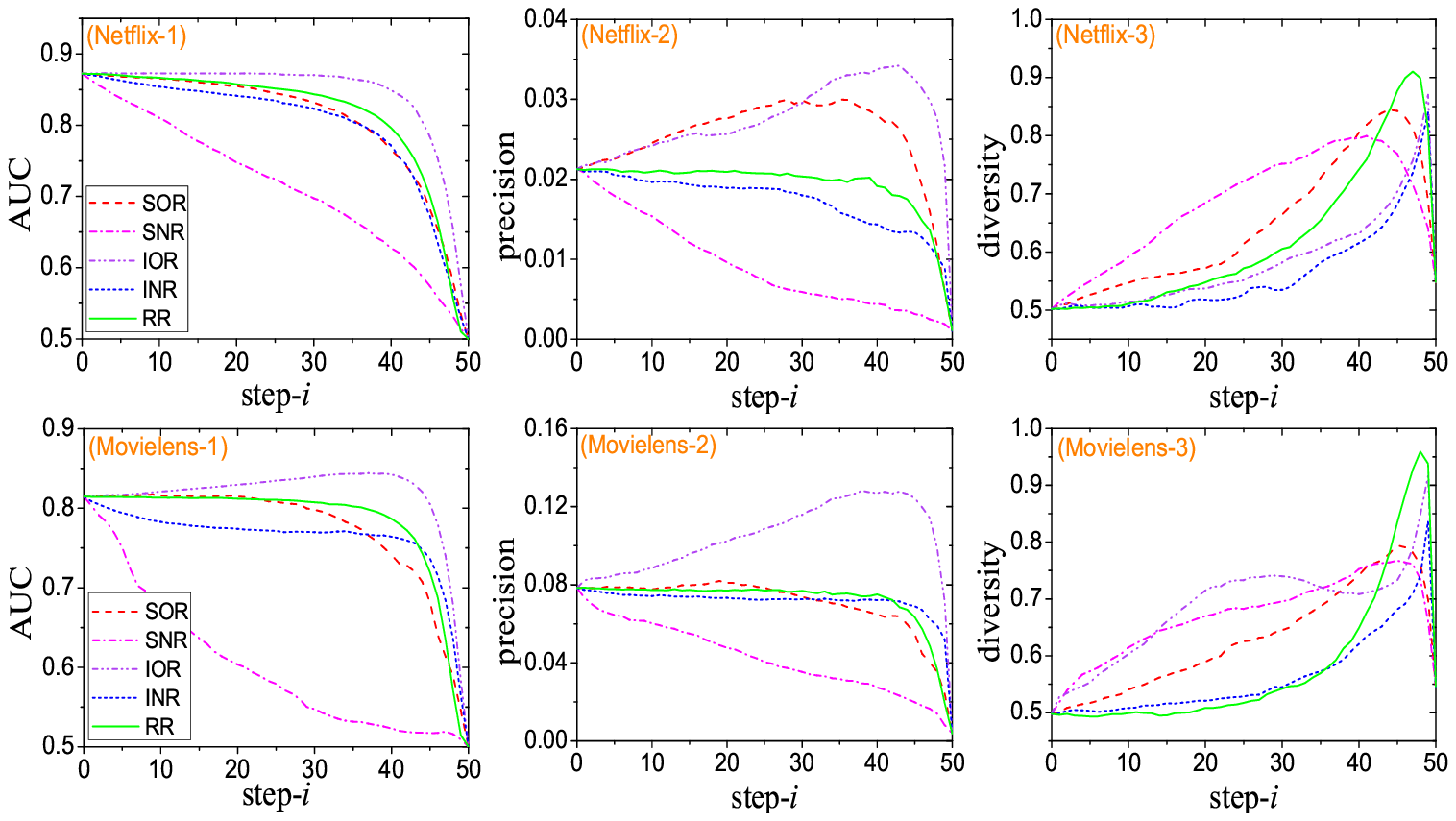}\\
  \caption{
  {\bf The variation tendencies of $AUC$, $P(L)$ and $H(L)$ with the macro-step increases.} step-$i$ is named the identifier of $i$th macro-step. The results of Netflix are shown in sub-figures (Netflix-1), (Netflix-2) and (Netflix-3), and those of Movielens are shown in sub-figures (Movielens-1), (Movielens-2) and (Movielens-3). This figure focuses on the time-aware algorithms.}\label{fig-Time-Sup}
\end{figure*}

\begin{figure*}[!ht]
  \centering
  \includegraphics[width=15cm]{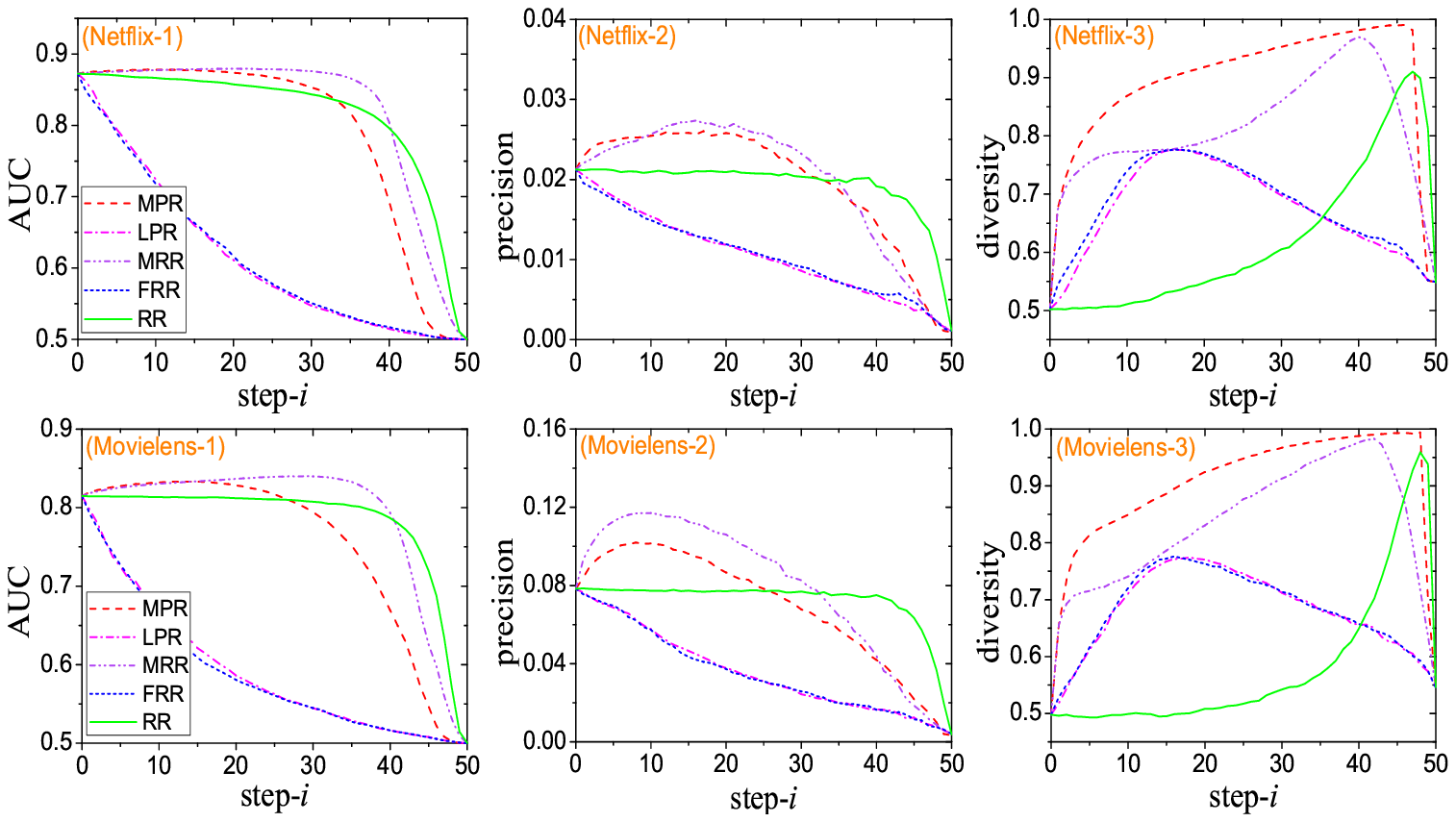}\\
  \caption{
  {\bf The variation tendencies of $AUC$, $P(L)$ and $H(L)$ with the macro-step increases.} step-$i$ is named the identifier of $i$th macro-step. The results of Netflix are shown in sub-figures (Netflix-1), (Netflix-2) and (Netflix-3), and those of Movielens are shown in sub-figures (Movielens-1), (Movielens-2) and (Movielens-3). This figure focuses on the topology-aware algorithms.}\label{fig-Topo-Sup}
\end{figure*}

\end{document}